\pagebreak
\documentclass[twoside]{sfm}

\input{sf.def}

\begin{document}

\thispagestyle{titlehead}

\setcounter{section}{0}
\setcounter{figure}{0}
\setcounter{table}{0}

%%%%%%%%%%%%%%%%%%%%%%%%%%%
% !!!
\markboth{Auri\`ere et al.}{Ap-star descendants }

\titl{Descendants of Magnetic and non-Magnetic A-type Stars}{Auri\`ere M.$^1$, Ligni\`eres F.$^1$, Konstantinova-Antova R.$^{2,1}$, Charbonnel C.$^{3,1}$, Petit P.$^1$, Tsvetkova S.$^2$, Wade G.$^4$}
{$^1$Institut de Recherche en Astrophysique et Plan\`etologie, Toulouse, France email: {\tt michel.auriere@irap.omp.eu} \\
 $^2$Institute of Astronomy and NAO, Sofia, Bulgaria\\
 $^3$Geneva Observatory, Geneva, Switzerland\\
 $^4$Royal Military College of Canada, Kingston, Canada}

\abstre{
We have studied the magnetic field of about 50  active and non-active single G-K-type red giants by means of spectropolarimetry with Narval and ESPaDOnS. 30 giants have been significantly Zeeman-detected. A close study of the 17 giants with known rotational periods shows that the measured magnetic field strength is correlated to the rotation, in particular to the Rossby number. 4 giants for which the magnetic field is measured to be outstandingly strong with respect to the rotational period or the evolutionary state are identified as probable Ap-star descendants. We detail their magnetic properties and propose criteria to identify Ap-star descendants.
}

\baselineskip 12pt

\section{Introduction}

A-type main sequence stars have no convective envelopes and are rather rapid rotators.
Leaving the main-sequence, a convective envelope is created which deepens, while the radius increases and the rotation slows. A dynamo-driven magnetic field is expected to be created in the case of the fastest non-magnetic or weakly magnetic A-type main sequence stars, and its strength should weaken when the rotation decreases. 
For  Ap-stars, the strength of the (fossil) magnetic field is expected to decrease as the radius decreases and an interaction between the existing magnetic field and the convective envelope should occur as well. Up to now the magnetic activity on the red giant branches was considered as limited to a small number of evolved giants that had been studied using indirect activity indicators, namely chromospheric or coronal emission, or photometric variations. With the introduction of new generation optical polarimeters, ESPaDOnS at CFHT or its twin Narval at TBL new sensitive magnetic surveys of evolved intermediate-mass stars became possible (Konstantinova-Antova et al. 2013 \cite{konst13}, Auri\`ere et al. in preparation). We will give an overview of our results concerning a sample of about 50 giants in Sect. 2,  emphasizing on the case of Ap-star descendants. In Sect. 3 we present our 4 Ap-star-descendant candidates and conclude in Sect. 4.

\section{Magnetic fields at the surface of active single G-K giants }%\label{nlte}

Using Narval at TBL and ESPaDOnS at CFHT (Donati et al. 2006 \cite{don06}), we have made a survey of about 50 single G-K giants chosen as favorable targets for Zeeman-detecting a surface magnetic field. Narval and ESPaDOnS are twin spectropolarimeters, each instrument consisting of a cassegrain polarimetric module connected by optical fibres to an echelle spectrometer. In polarimetric mode, the instrument simultaneously acquires two orthogonally-polarized spectra covering the spectral range from 370 nm to 1000 nm in a single exposure, with a resolving power of about 65000. To obtain a high-precision diagnosis of the spectral line circular polarization, least-squares deconvolution (LSD, Donati et al. 1997 \cite{don97}) was applied to each reduced Stokes $I$ and $V$ spectrum. From these mean Stokes profiles we computed the surface-averaged longitudinal magnetic field  $B_\ell$ in G, using the first-order moment method adapted to LSD profiles (Donati et al. 1997). The procedure employed is the same as described for example by Auri\`ere et al. 2012 \cite{aur12} .

From our survey we have Zeeman-detected 30 stars from our red giant sample.  24 upon the 25 giants presenting signs of activity are Zeeman-detected, as well as 6 upon 18 bright giants enabling to reach deep magnetic sensitivity. Measured $B_\ell$ span from a few tenths of G to about 100 G. $|B_l|_{max}$ is weaker than 10 G for 70 \% of the Zeeman-detected giants. Figure 1 shows the stars of our sample in the Hertzsprung Russell diagram. Tracks of the standard evolutionary models of Charbonnel and Lagarde (2010 \cite{charb10}) are shown . The Zeeman-detected stars are represented with filled circles, those which are not detected correspond to empty circles. The magnetic stars are found in general in the first dredge up zone or at the core He-burning phase (Auri\`ere et al. in preparation).

\begin{figure}[!t]
\begin{center}
 \includegraphics[width=7.5cm]{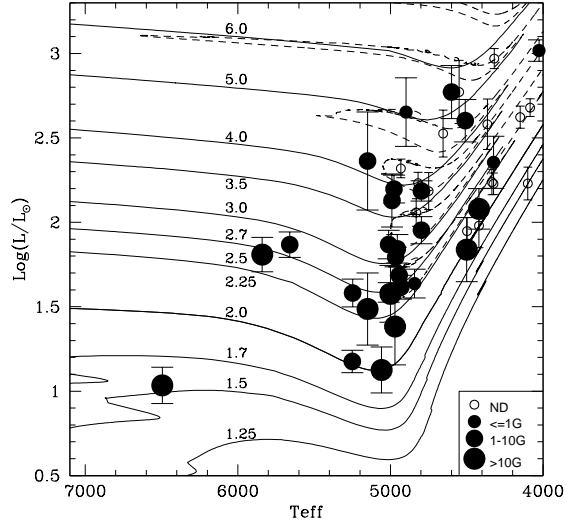}
\vspace{-5mm}
\caption[]{Position of the Red Giants of our sample in the Hertzsprung-Russell Diagram. Tracks of the standard evolutionary models of Charbonnel and Lagarde (2010 \cite{charb10}) are shown.}
%\label{ca1_ca2}
\end{center}
\end{figure}

17 of the observed giants have a determined rotational period $P_{rot}$. Evolutionary models with rotation  of Charbonnel \& Lagarde (2010 \cite{charb10}) were used to compute the convective turnover time $\tau_{conv}$ at the height of $H_p/2$ above the base of the convective enveloppe. Using the  $P_{rot}$ values and these $\tau_{conv}$ , one can obtain their ratio which is the Rossby number $R_o$. $R_o$ measures the efficiency of the solar-type dynamo. Figure 2 shows that there exist an exponential relation between the magnetic field strength (taken as $|B_l|_{max}$) and $R_o$. This suggests that the origin of the magnetic field is the same for all the stars fulfilling the relation, and since $R_o$ is smaller than 1.5 and even reaches 0.08, a solar-type dynamo could be efficient. Figure 3 also shows a rather tight correlation between the $S$-index (measuring the chromospheric Ca II H\&K lines) and $|B_l|_{max}$. As will be discussed in the following sections, there are also outliers to these relations: some giants corresponding to over-strong fields will be presented as possible Ap-star-descendant candidates. In this case the departure from the relation presented in Fig. 3 can be explained easily. In the case of dynamo-driven magnetic fields, both small and large magnetic scales are expected to be present: both scales contribute to chromospheric emission, while the small scales which are with opposite polarities can cancel each other, reducing the observed averaged longitudinal field. On the other hand, in the case of Ap-star-descendant fossil fields, the large scale of a dipole dominates, which favors the $B_\ell$ measurements.

\begin{figure}[!t]
\begin{center}
 \includegraphics[width=7.5cm]{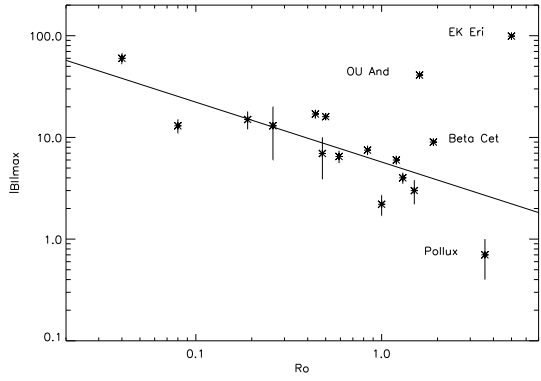}
\vspace{-5mm}
\caption[]{Variations of the strength of the magnetic field ($|B_l|_{max}$) with the Rossby number.}
%\label{ca1_ca2}
\end{center}
\end{figure}

\begin{figure}[!t]
\begin{center}
 \includegraphics[width=7.5cm]{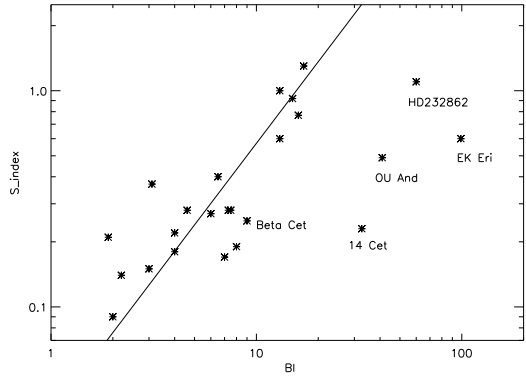}
\vspace{-5mm}
\caption[]{Variations of the $S$-index with the strength of the magnetic field ($|B_l|_{max}$) .}
%\label{ca1_ca2}
\end{center}
\end{figure}

\section{4 Ap-star-descendants candidates}

In our investigation, we have identified 3 very active giants whose $|B_l|_{max}$ is stronger than predicted from their rotational period, and one whose $|B_l|_{max}$ is stronger than expected from its evolutionary status. A close study leads us to propose that these giants are convincing Ap-star-descendant candidates.

\subsection{EK Eridani}

EK Eridani (HD 27536, HR 1362) has been proposed as being an Ap-star descendant by Stepie\'n (1993) \cite{sti93}: because of its long rotational period of 308.8 d. (Dall et al. 2010 \cite{dall10}),  its strong activity is unlikely to be driven by a solar-type dynamo. Our Zeeman-study (Auri\`ere et al. 2008 \cite{aur08}, 2011 \cite{aur11}) has unveiled a strong surface magnetic field with dipolar topology and turned the Ap-descendant hypothesis from speculation to a convincing hypothesis. The properties of EK Eri which support the hypothesis of an Ap-star descendant are: - A large $R_o$ and a $|B_l|_{max}$ well above the value expected according to the relation shown Fig. 2. - The departure from the relation $S$-index versus $|B_l|_{max}$ presented in Fig. 3. - A dominant dipolar field. - The correlation between $B_l$ and the photometric modulation. Since the photometric modulation has been relatively stable during the 30 past years, the magnetic field topology should have been so as well, which can be explained by the presence of the remnant of the Ap-star ''fossil" field. EK Eri can be therefore considered as the archetype of the Ap-star descendants.

\subsection{OU Andromedae}

OU And (HD 223460, HR 9024) is a well known very active single giant (e.g. Gondoin 2003 \cite{gond03}) crossing the Hertzsprung gap (Auri\`ere et al. in preparation) with a $P_{rot}$ of 24.2 d (Strassmeier et al. 1999 \cite{strass99}). We detected at its surface a strong magnetic field ($|B_l|_{max}$ = 41.4 $\pm$ 1.5 G) which causes the star to deviate significantly from the relations $|B_l|_{max}$ versus $R_o$  and  $S$-index versus $|B_l|_{max}$ presented in Fig. 2 and Fig. 3 respectively. It was also remarked that this star is almost at the same evolutionary state and level of activity in the classical activity indicators as 31 Com (Strassmeier et al. 2010 \cite{strass10}) which is a fast rotator (Auri\`ere et al. in preparation).  For these reasons we consider that the origin of the strong magnetic field of OU And is from an Ap-star progenitor, whereas it is from a solar-type dynamo in 31 Com.

\subsection{$\beta$ Ceti}

$\beta$ Ceti (HD 4128, HR 188) is also a well known very active single giant (e.g. Ayres et al. 2001 \cite{ayr01}). Tsvetkova et al. (2013 \cite{tsv13}) have Zeeman-detected and studied its magnetic field, and they confirmed that $\beta$ Ceti is a giant in the central He-burning phase. They also found that its $P_{rot}$ is of 215 d., well above what was expected from its measured strong $S$-index (Young et al. 1989 \cite{young89}). Using this $P_{rot}$ and its derived $R_o$, $\beta$ Ceti is an outlier from the $|B_l|_{max}$ versus $R_o$ relation presented in Fig. 2. At the end, since they find that the topology of the surface magnetic field is dominated by a dipole, 
Tsvetkova et al. (2013 \cite{tsv13}) concluded that $\beta$ Ceti was very likely an Ap-star descendant.

\subsection{14 Ceti}

14 Ceti (HD 3229, HR 143) is a subgiant F5IV for which a rather strong magnetic field (about 30 G) was detected (Auri\`ere et al. 2012 \cite{aur12}). 14 Ceti is the only Zeeman-detected star between the F0p magnetic stars and the F7 solar-type stars. Though we made an intense survey of its magnetic field (Auri\`ere er al. 2012 \cite{aur12}) we could not determined any period in the Stokes V signal and derived that $P_{rot}$ is very likely greater than a few months. Since 14 Ceti is just leaving the main sequence, its external convective enveloppe is very thin and the $\tau_{conv}$ computed at the bottom of the CE is only about 4 days when using the models of Charbonnel and Lagarde (2010). Therefore only a very fast rotator (observed almost  pole-on to be consistent with the small observed $v\sin i$ $<$ 5 km s$^{-1}$) would be able to provide the strong observed magnetic field if driven by a solar-type dynamo. Finally, the Ap-star descendant hypothesis appears more likely for explaining the activity properties of 14 Ceti (Auri\`ere er al. 2012 \cite{aur12}).

\section{Conclusions}

Our investigation of the surface magnetic field of active single giant stars has shown that the most active of them have common properties which suggest that their magnetic activity could be in great part driven by a dynamo of the solar-type (Auri\`ere et al. in preparation). Some outliers of the relations presented on Fig. 2 and Fig. 3, may be peculiar cases: for example the very fast rotator HD 232862 could be of FK Com type, and Pollux has a very weak magnetic field which is certainly not driven by a solar-type dynamo. 

In addition we have identified some active giants which are outliers of the common behaviour and could be Ap-star descendants, with  their masses being in the relevant range (Power et al. 2007 \cite{power07}). Their characteristics are the following: - A strong  $|B_l|_{max}$ with respect to their rotation rate or/and their evolutionary state (e.g. Fig.2) - Significant deviation from the $S$-index versus $|B_l|_{max}$ relation (Fig. 3). - A simple magnetic topology (inclined dipole). - Long term stability of the periodic behaviour ($B_l$ / Radial Velocity / Activity).

Within these criteria (only EK Eri fulfil all of them), 4 active single giants are proposed as being probable Ap-star descendants. They are located at different stages of the evolution on the RG branches: 14 Ceti (1.5~$M_{\odot}$) is just leaving the main sequence and entering the Hertzsprung gap; OU And (2.5~$M_{\odot}$) is in the Hertzsprung gap; EK Eri (1.9~$M_{\odot}$) is at the base of the RGB ; $\beta$ Ceti (3.5~$M_{\odot}$) is in the central He-burning phase. For all these stars, the hypothesis of the magnetic flux conservation as the radius increases provides reasonable values of magnetic strength for the Ap-star progenitors.

\bigskip
{\it Acknowledgements.} The authors thank the TBL and CFHT teams for providing service observing with Narval and ESPaDOnS. Observing time at TBL was founded by PNPS of CNRS, OPTICON network and Bulgarian NSF.

\end{document}